\documentclass{elsart}
\usepackage{natbib}
\begin{document}
\runauthor{Cicero, Caesar and Vergil}
\begin{frontmatter}
\title{New Generation Atmospheric Cherenkov Detectors}
\author{Frank Krennrich}

\address{ Physics \& Astronomy Department, Iowa State University,\\
            Ames, IA, 50011, USA  }

\begin{abstract}
High energy $\gamma$-ray astronomy has been established during
the last decade through the launch of the Compton Gamma Ray 
Observatory (CGRO) and the success of its ground-based counterpart, 
the imaging atmospheric Cherenkov technique. 
In the aftermath of their important and surprising scientific results a
worldwide effort developing and designing new generation atmospheric
Cherenkov detectors is underway.  These novel instruments will have
higher sensitivity at E $\rm > $ 250 GeV, but most importantly, will
be able to  close the unexplored energy gap between 20 GeV and 250 GeV.
Several ground-based detectors are proposed or under construction. 
Aspects of the techniques used and sensitivity  are discussed  in 
this overview paper.   The instruments cover largely complementary 
energy ranges and together are expected to explore the $\gamma$-ray 
sky between 20 GeV and 100 TeV with unprecedented sensitivity.
\end{abstract}

\begin{keyword} $\gamma$-ray astronomy: GeV - TeV energies --
ground-based instruments: atmospheric Cherenkov detectors  
\end{keyword}
\end{frontmatter}

\section{Introduction}

The EGRET detector on-board of CGRO has stimulated the field
of high energy astrophysics by the detection of 271 $\gamma$-ray 
sources \cite{Hartmann99} in the energy range between 10 MeV - 20 GeV.   
In parallel, the success of the ground-based imaging atmospheric Cherenkov
technique, operating between 250~GeV - 50~TeV, has demonstrated that 
$\gamma$-ray astronomy can be expanded well beyond GeV energies.   
A pioneering experiment, using the Whipple Observatory 10~m imaging 
telescope, achieved the first unequivocal detection of the 
Crab Nebula~\cite{Weekes89}, a plerion or pulsar-powered synchrotron nebula.  
More of a surprise was the discovery of TeV emission from a subclass 
of active galactic nuclei (AGNs), the so-called blazars:  Mrk 421~\cite{Punch92},  
Mrk 501~\cite{Quinn96}  and 1ES~2344+514~\cite{Catanese98}. 
Although EGRET has reported 66 high-confidence (and 27 low-confidence)
blazar identifications with redshifts between z~=~0.0018 - 2.286,  the
blazars at TeV energies are all nearby with redshifts between z~=~0.031 - 0.044.
It is tempting to speculate~\cite{Stecker97} that TeV $\gamma$-rays
are absorbed by the extragalactic IR background when traveling distances 
comparable to most EGRET blazars.
However, whether or not the non-detection of most EGRET blazars at TeV 
energies can be attributed to their interaction with IR background photons 
($\gamma\gamma \to e^+e^-$), or is simply due to a spectral break at the source,
remains an open question,  because of the lack of data above
20 GeV.

Contrary to earlier predictions~\cite{Stecker93}  for the IR 
background density, recent TeV $\gamma$-ray observations show that the
energy spectra of Mrk 421 and Mrk 501 extend beyond 
10 TeV~\cite{Aharon97,Samuelson98,Krennrich99}, 
the highest energies detected from any AGN. This is still consistent with
detailed models~\cite{Primack98} for the IR density.  Apart from constraining
IR background models, the high energy $\gamma$-rays from nearby blazars
can also provide important data to test $\gamma$-ray production models.
In particular the possibility of acceleration of hadronic cosmic rays in 
jets of blazars can be studied.  Extending the measurements to 100 TeV
and down to 20 GeV could provide the crucial evidence for the understanding
of particle acceleration in jets.  Also the fact that the $\gamma$-ray  peak 
emission  occurs at vastly different energies, e.g., at a few GeV for 
3C 279~\cite{Wehrle98} and at a few hundred GeV for Mrk 501~\cite{Catanese97} 
(a weak detection by EGRET has been reported just recently~\cite{Kataoka99})  
emphasizes 
the importance of a big energy coverage in blazar studies.     A  wide range 
of energies from 20~GeV up to 100~TeV with a good energy resolution is desirable
to study spectral features. 

The interest in galactic $\gamma$-ray astronomy was ignited by the detection 
of two additional plerions PSR 1706-44~\cite{Kifune95} and 
Vela~\cite{Yoshikoshi98} and even more so, by a detection of a 
shell-type supernova remnant, SN 1006~\cite{Tanimori98a} by the CANGAROO 
telescope.  One of the primary motivations for galactic $\gamma$-ray astronomy
is the understanding of supernova shock acceleration and the origin of the
galactic cosmic rays.  The search for their most promising acceleration sites,
in shell-type supernova remnants~\cite{Drury94},  requires a good sensitivity for
extended~\cite{Buckley98} $\gamma$-ray emission.  Future atmospheric Cherenkov detectors will
also address the sensitivity for extended sources like the galactic plane 
and shell-type supernova remnants.

GeV and TeV $\gamma$-ray measurements have caused a wide scientific interest.
It is crucial for the understanding of the $\gamma$-ray emission 
processes to close the energy gap between 20~GeV - 250~GeV and to advance 
the sensitivity between 250~GeV - 100~TeV. 
The gap between 20~GeV and 250~GeV can be explored with future satellite 
and ground-based instruments.   Both techniques are currently beeing
investigated:   the satellite-based GLAST~\cite{Glast96} detector with a 
relatively small 
collection area but a wide field of view,  and the ground-based atmospheric 
Cherenkov technique with a small field of view but a large collection area. 
The energies above 250~GeV will remain the domain of the ground-based 
detectors.  In this overview of atmospheric Cherenkov detectors, the 
detection  principle and their anticipated sensitivity, energy range and 
angular resolution are discussed.  In Section 2 important design 
considerations for atmospheric Cherenkov detectors are briefly outlined.  
In Section 3 the individual detectors are described. Section 4 gives a 
summary of the anticipated performance of the various instruments 
emphasizing their individual strengths.

\section{Design Considerations}
The design of a new  instrument  is driven by science.   From 
the previous section it is evident that the need for a wide range in energy 
(20~GeV - 100~TeV), wide field-of-view and   spectroscopic capabilities 
puts many requirements on future detectors. The detection principle of 
$\gamma$-rays above 20~GeV from the ground
is based on the measurement of Cherenkov light from the electromagnetic
atmospheric cascade initiated by the $\gamma$-ray primary.  The Cherenkov
light emitted from the secondary $\rm e^\pm$ over a range of altitudes 
(6 - 20 km atmospheric height) is focused onto an area of 200 - 300~m 
diameter at ground defining the light pool.  The collection area, the 
area from which a shower can be detected,  is in the order of 
 70,000~$\rm m^2$, although the area also depends on the energy and 
detailed detector design.    The intrinsically large collection area 
of atmospheric Cherenkov detectors (order of $\rm 10^5$ 
larger than EGRET's collection area) would ideally extend the sensitivity 
above 20~GeV, where EGRET's measurements are limited by statistics.
The large collection area also provides the means for the
detection of low fluxes in the 1~TeV - 100~TeV regime with a good sensitivity.  
In particular large zenith angle observations, which provide an even
larger collection area (several 100,000~$\rm m^2$), is efficient for 
the detection of the highest energies 
above 10~TeV~\cite{Sommers87, Krennrich97, Tanimori98b}.
The following issues need to be addressed by future detectors to reach the 
objectives outlined in the previous section:  a low energy threshold of 
20~GeV, improved sensitivity between 200~GeV - 100~TeV,  high angular 
resolution (few arcminutes) and good energy resolution (10\%).

\subsection{Low Energy Threshold}

The detection of low energy $\gamma$-ray air showers requires
triggering on faint Cherenkov light flashes, e.g., 2.5 photons per $\rm m^2$
from a 50 GeV $\gamma$-ray shower\footnote{This value comes from Par\'e 1996
and is valid for the Th\'emis site. The absolute value depends on the 
altitude and atmospheric conditions at the observational site.}.     
A limitation arises from the night 
sky background light (NSB $\rm \approx 2 - 4 \times 10^{12} \: photons /m^2/s/sr $) 
through its fluctuations $\rm \sqrt{NSB}$:  however, since the Cherenkov 
pulses are extremely short (5 - 10 ns),  the $\rm NSB$  can be greatly reduced 
(e.g., 0.8 -1.6  $\rm photons/m^2$ within 5 ns for a photosensitive detector  
with $\rm 0.6^{\circ}$ 
sensitive diameter\footnote{The  minimum solid angle acceptance required for an atmospheric 
Cherenkov detector is $\rm \approx  0.6^{\circ}$ to cover  the angular 
extend over   which the Cherenkov light from a sub-TeV $\gamma$-ray shower from 
a point source occurs determined by its intrinsic angular size and shower height 
fluctuations.     However,  the light distribution of a  $\gamma$-ray  
shower image is elliptical and peaked and can be
described by its width and length with a scale of $\rm 0.15^{\circ} 
\times 0.3^{\circ}$.    With  imaging  telescopes the aperture for the 
triggering pixels can be reduced through the use of a fine pixellation
camera, effectively reducing the $\rm NSB$.}).
Air showers can only be detected if the Cherenkov signal exceeds 
several times $\sqrt{\rm NSB}$.  This can be quantified by the signal to noise ratio, 
defined as the number of Cherenkov photons over the night sky fluctuation.   
A low energy threshold detector requires the optimization of the signal to 
noise ratio, e.g., by minimizing the solid angle acceptance $\rm \Omega$ of 
the triggering photodetectors or by shortening  the coincidence time window 
for the Cherenkov  pulses in different  photodetectors effectively reducing 
the $ \sqrt{\rm NSB}$ contribution.  
For imaging telescopes the solid angle acceptance covered by the 
minimum required photodetectors (typically photomultipliers)    
should be comparable to the $\gamma$-ray image size scale. The time window 
$\rm \tau$ should be close to the intrinsic pulse width of the Cherenkov flash.

 In order to lower the energy threshold of  atmospheric Cherenkov detectors  
(presently E $\rm > 250 \: GeV$) to 50 GeV (20 GeV), the number of Cherenkov 
photons collected has to be substantially increased, e.g. through increasing 
the mirror area $\rm A_{mirror}$.  The energy threshold scales as 
$\rm E_{thres} \propto  \sqrt{\rm NSB\: \Omega \: \tau/A_{mirror}}$.  
Simple extrapolation from existing instruments, for example the Whipple 10~m 
 telescope which operates at a high signal to noise ratio, suggests 
a mirror area of $\rm \sim 1800~m^2$ and $\rm \sim 11,700~m^2$ to reach 50 GeV 
and 20 GeV respectively.
However, it is important to notice, that the overall quantum efficiency of  
existing detectors (the convolution of mirror reflectivity and photomultiplier 
quantum efficiency) does not exceed $\approx$ 10\%~\cite{Mirzoyan94}, dominantly 
hampered by the low quantum efficiency of photomultiplier tubes.   Improving the 
light collection efficiency also lowers the energy threshold.

Future atmospheric Cherenkov detectors described in Section 3 use various 
means to achieve a lower energy threshold.
Arrays of 10~m imaging telescopes, proposed by the VERITAS~\cite{Weekes97} 
and HESS~\cite{Hofmann97} collaboration,  aiming for an energy threshold 
of $\rm \sim 100$~GeV, are based on a moderate increase of mirror area, combined 
with fast electronics and fine pixellation imaging cameras.
The MAGIC~\cite{Barrio98} collaboration proposes a single 17~m imaging telescope aiming for 
a  low energy threshold through a modestly  increased mirror area together 
with high quantum efficiency phototubes.     A dramatic increase of mirror 
area is  currently explored by the STACEE~\cite{Ong97} and CELESTE~\cite{Pare97}
collaborations,  where 48 - 160 heliostats of solar power 
plants give a gigantic mirror area for detection  of $\gamma$-rays down to $\approx$ 30 GeV.   

\subsection{Sensitivity, Angular and Energy Resolution}

The rejection of much more numerous cosmic-ray induced showers
has been pivotal in establishing the sensitivity of ground-based 
TeV astronomy~\cite{Weekes89}.    In constrast to satellite instruments, 
which utilize anti-coincidence scintillator shields inhibiting the detector from 
triggering on charged cosmic-ray particles,  ground-based instruments rely 
mostly\footnote{However, at the hardware trigger level some background
is rejected, because at E~$\rm \approx$~100 GeV (300 GeV) a $\gamma$-ray induced shower produces
10 times (4 times) more light than a proton induced shower of the same energy. Also, 
the field-of-view determines the background level.  Therefore,
 a comparison of instruments should include both, the off-line and the hardware rejection
capability. } on background rejection in off-line analysis.
The imaging technique has provided an efficient means to separate $\gamma$-ray 
from hadronic initiated showers which can be expressed by a 
quality factor Q defined as $\rm Q = \varepsilon/ \sqrt{\kappa}$ with 
$\rm \varepsilon$ = efficiency for $\gamma$-rays and $\rm \kappa$ 
= efficiency for cosmic-ray events. 
Because  the detection of $\gamma$-rays  is background  dominated 
($\rm  \sim 10^3$ more cosmic rays than $\gamma$-rays), the 
sensitivity of any atmospheric Cherenkov detector depends critically on its 
background rejection capability.   

The measurement of the arrival direction of $\gamma$-rays depends on the 
detection technique:  an arrival time measurement of
the wavefront of Cherenkov photons (as used in STACEE and CELESTE)  or by 
the analysis of the shower image orientation in imaging telescopes 
(VERITAS, HESS and MAGIC).   A good angular 
resolution is particularly important for point-source sensitivity at energies 
below 100 GeV, where isotropically arriving cosmic electrons constitute an
additional,  non-hadronic background, which cannot be rejected through $\gamma$/hadron separation.   
The angular resolution naturally improves with the 
$\gamma$-ray primary energy from $\rm 0.1^{\circ}$ below 
100 GeV up to the  $\rm 0.02^{\circ}$ range at TeV energies.

The energy resolution is important for extracting the physics of
the $\gamma$-ray source.  For example the sharp spectral
breaks in pulsars  as suggested by the polar cap model~\cite{Daugherty96} and generally the 
measurement  of the spectral breaks of most EGRET sources between 20 GeV 
and 200 GeV provides motivation for an  energy resolution in the range of 10\%.

\section{Future Projects}

\subsection{Imaging Telescopes}

The imaging technique uses optical reflectors
(e.g. Whipple 10~m telescope) with a tessellated mirror structure 
and a matrix of fast photomultipiers in the focal plane. With this 
configuration an image of the Cherenkov light of an air shower is
measured and analyzed.   Weekes et al.~\cite{Weekes89} have demonstrated
that the analysis of image shape and orientation is very efficient 
in distinguishing $\gamma$-ray from cosmic-ray initiated air showers.
A Q-factor of $\rm \approx$ 10 (99.7\% of cosmic rays are rejected while 
keeping 60\% $\gamma$-rays) with the Whipple 10~m telescope has been pivotal in  
establishing the imaging technique, which is  to date the only technique which 
has detected $\gamma$-ray sources above 250 GeV at a level of $\rm \geq 10 \sigma$.
The Crab Nebula is detected at a rate of 2 $\gamma$-rays per minute with a 
sensitivity of about 7$\rm \sigma$ per hour. Based on this success, two different 
concepts have been proposed to increase the sensitivity further:  
the development of an optimized single large telescope (MAGIC) and stereoscopic 
imaging using multiple telescopes (VERITAS and HESS).

\subsubsection{Single Telescope Imaging: MAGIC}

The potential of improving the single telescope imaging method has been 
recognized by several groups: CAT, MAGIC and the Whipple collaboration. 
Although the CAT imaging telescope employs a relatively small 18 $\rm m^2$ 
mirror (75 $\rm m^2$ for the Whipple 10~m) an 
energy threshold of 200 GeV has been reached~\cite{Goret97}.  This is 
due to fast electronics and a fine pixellation camera ($\rm 0.12^{\circ}$
vs. $\rm 0.25^{\circ}$ for the Whipple 10~m) optimizing the signal to
noise ratio at the trigger level.  The combination of a 10~m telescope with 
a pixellation of $\rm 0.12^{\circ}$ is currently 
pursued by the Whipple collaboration (GRANITE III) by upgrading the 10~m 
telescope~\cite{Lamb95}.  An energy threshold of 120 GeV is anticipated.
 
To push this strategy further, the MAGIC collaboration pursues the design 
of a 17~m diameter telescope.   The  concept  is  based  on 
increasing the mirror area, a better quantum efficiency 
(45\% GaAsP photocathode) of the photon detection devices and fast speed 
electronics.     Estimates given by 
the MAGIC collaboration quote an energy threshold of 30 - 40 GeV
using standard photomultipliers and 15 GeV using photodetectors with GaAsP 
photocathodes~\cite{Barrio98}.  Simulations by the MAGIC
group show, that 20 GeV $\gamma$-ray showers produce images which contain 
good information about the arrival direction suggesting an angular resolution of
$\rm 0.2^{\circ}$ near threshold\footnote{Note that the angular resolution 
is a function of the primary energy and improves substantially at higher 
energies.}, and a good   background rejection\footnote{Cosmic-ray background
from hadrons, muons and electrons has been considered.}  with a Q-factor of 
$\rm 6$.   An energy resolution of 50\% at the threshold energy and 20\% at 
100~GeV is quoted.

\subsubsection{Stereoscopic Imaging: VERITAS and HESS}

A logical extension of the single telescope imaging technique is the
stereoscopic detection of air showers with multiple instruments which has 
been first demonstrated by Grindlay~\cite{Grindlay72}.     The detection 
of a $\gamma$-ray signal with multiple imaging telescopes was demonstrated 
by Daum et al.~\cite{Daum97}  and Krennrich et al.~\cite{Krennrich98}.  
Impressive results have 
come from the HEGRA telescope array (4 telescopes of 8.5 $\rm m^2$ mirror area each) using 
relatively small reflectors and a pixellation of $\rm 0.25^{\circ}$, showing 
a good angular resolution and excellent background rejection at 
1~TeV~\cite{Konopelko98}.
Two next generation multi-telescope projects are under development; 
the VERITAS array (7 $\rm \times$ 10~m telescopes) in the northern hemisphere  
(Arizona) and the HESS project ($\rm 16 \times 10~m$ 
telescopes) in the  southern  hemisphere (Namibia).
The major objective of those multi-telescope installations is the stereoscopic
detection of $\gamma$-ray sources above 100~GeV with a high  sensitivity, 
angular resolution and energy resolution~\cite{Aharonian97b, Aharonian97c}.

The multi-telescope imaging technique is based on the stereoscopic 
view of $\rm \gamma$-ray showers (Figure 1, 2).  This provides an angular resolution of  
$\rm 0.08^{\circ}$  at  100~GeV     and $\rm 0.02^{\circ}$
for the highest energies for a VERITAS type detector~\cite{Vassiliev98},  which  
by itself improves
the background suppression of the cosmic-ray induced showers in comparison to a
single telescope.  In addition, the image shapes of air showers can be 
better constrained with several telescopes and be reconstructed in 3-dimensional 
space providing a measurement of the height of shower maximum,  shower impact
point on the ground (Figure 2) and the light density at different locations within the 
Cherenkov light pool.    As a result,   the $\gamma$-ray energy~\cite{Carterlewis98} can 
be better measured with a resolution of    13\% - 18\% (corresponding to 10 
TeV and 100 GeV).           Also the Q-factor improves through a better 
classification of $\gamma$-ray, cosmic-ray 
or muon images and excellent angular resolution\footnote{Note for a point source
the background rejection is due to two different factors, the angular resolution and
the distinction of $\gamma$-rays from hadronic showers and single muons.}.  
Also, the collection area for the stereoscopic operation of a 7-telescope array requiring
a 3-telescope coincidence is increased to 200,000~$\rm m^2$. 
 Monte Carlo simulations suggest that the point-source sensitivity of the VERITAS 
array, e.g., at 300 GeV~\cite{Vassiliev98}, is a factor of 10 better than 
with the currently operating Whipple 10~m telescope.  

The energy threshold (analysis threshold) of arrays can be lower than for an
individual telescope.
First,   the rejection of local muons (a muon can be detected up to 80 m 
distant from telescope) through an array trigger will be important at $\gamma$-ray 
energies below 300 GeV, where they constitute a major background.   Remaining muons
falling between the telescopes can be rejected by their parallactic 
displacement\footnote{As opposed to distant $\gamma$-ray showers, local muons 
show a parallactic offset when comparing images in different telescopes.}. 
Lastly, faint Cherenkov flashes (barely triggering) do not produce 
a well defined image shape and hence not much information about the
nature of the primary is available.  Those images are usually rejected 
in the single telescope analysis. 
However,  using the stereoscopic view~\cite{Krennrich95, Hillas96}  
$\gamma$-ray showers differ from hadronic showers: images from hadronic primaries
 show a more irregular parallactic displacement than $\gamma$-ray induced showers.  
This method can be used to provide hadronic background rejection at energies close 
to the trigger threshold of telescope arrays.   For VERITAS a trigger and analysis
threshold of 50 GeV seems possible using stereoscopic reconstruction methods, 
and the limit arises mostly from the night sky background fluctuations.


\subsection{Light Pool Sampling with Heliostats}

The potential of utilizing solar power plants for $\gamma$-ray 
astronomy has been recognized~\cite{Tumer91}, because  mirror
areas of several thousand square meters provide the 
necessary signal to noise ratio to trigger on E $>$ 20 - 300 GeV
$\gamma$-ray primaries.   The exploration of the lowest energies 
E $\rm \approx$ 20 GeV with a low cost device is the primary objective 
of the STACEE, CELESTE and GRAAL~\cite{Plaga95} projects.

The principle of detecting $\gamma$-rays with heliostats is shown in Figure 3.   
The Cherenkov light from an extensive air shower is collected
with steerable mirrors and reflected onto a stationary secondary mirror located
on the central tower.  Because the secondary mirror forms an image of the
locations of the heliostats it projects the light from each individual
heliostat onto a different position in the focal plane.  Photomultipliers
are used to detect and sample the light distribution. 
Due to the different times of flight between  different heliostats and the secondary 
mirror it is necessary to delay the signals with respect to each other and combine them 
afterwards into one trigger.            By forming an analog sum of 
the signals between several phototubes, the total amount of light
collected by all mirrors can be combined almost as if it were detected by a 
single large mirror\footnote{Alternatively, they can be combined in a digital 
sum after they have individually passed a discriminator.}, therefore 
providing a low energy threshold.
Because of the diameter of the Cherenkov light-pool of $\gamma$-ray induced showers at 
20 GeV (200 m diameter), the number of heliostats which participate in the trigger have 
to be limited to mirrors which fall into an area of that size.

The Cherenkov light distribution is sampled at different positions on the ground.
This provides information about the Cherenkov light intensity as a function of 
the position within the light pool.  Hadronic showers show more irregular azimuthal 
variations in the light density within the light pool than $\gamma$-ray showers,  
and this property can  be utilized to reject hadronic cosmic-ray showers.

The arrival direction is somewhat preset by a fairly narrow field of view
or solid angle acceptance of the configuration (angular extend $\rm 1.2^{\circ}$)\footnote{The
solid angle acceptance of individual heliostats is $\rm \approx 0.7^{\circ}$}.
The arrival direction can be measured by deriving the orientation of the shower 
wavefront from the delays between the signals from the different heliostats.    
The arrival direction reconstruction also requires information about the shower
core location which can be achieved by sampling the light density on the ground~\cite{Pare96} 
providing an angular resolution of $\rm 0.2^{\circ}$ with 40 heliostats (CELESTE).   
First light by the CELESTE collaboration has resulted in a tentative detection
of the Crab Nebula at an analysis threshold of 80 GeV~\cite{Smith98}.  


\section{Discussion}

The different techniques to detect $\gamma$-rays from the ground 
are largely complementary as emphasized in Figure 4.   The lowest energies
are explored by solar power plants starting at 20 GeV with a 
sensitivity for point sources.  Those instruments will operate
up to a few hundred GeV where their narrow aperture limits the 
detection of higher energy $\gamma$-rays.   The MAGIC project
targets a 15 - 40 GeV energy threshold exploring the  imaging
technique at the lowest energies.  Because of its $\rm 2.5^{\circ}-3.5^{\circ}$ 
field of view,  MAGIC could also provide a sensitivity for extended sources
and $\gamma$-ray burst counterpart searches. 
At higher energies of E $\rm >$ 100 GeV (possibly  50 GeV), VERITAS and HESS
can detect $\gamma$-rays over a big dynamic range of up to 100 TeV, whereas their
primary sensitivity is between 100 GeV - 10 TeV.  The high angular resolution and 
strong background suppression provides excellent sensitivity in the order of
a few milli-Crab.   Also, the combined field-of-view of VERITAS (or HESS) can
create maps of extended regions in the sky covering $\rm 10^{\circ}$ in diameter
with a single exposure.
An all-sky survey of the TeV sky will be carried out with MILAGRO~\cite{Yodh96}, 
also providing potentially important information where to point atmospheric Cherenkov
detectors.

From Figure 4 it becomes clear that the 20 - 200 GeV window will be
opened up for high  energy $\gamma$-ray astronomy by the alliance of
space-based (GLAST) and ground-based Cherenkov detectors.

\section*{Acknowledgements}
This work is supported by a grant from the U.S. Department of Energy. I am 
grateful to my colleagues who have provided me with detailed information about 
the status of experiments in particular W. Hofmann, E. Lorenz, 
R. Ong, M. Panter, G. Sinnis, D. Smith, S. Westerhoff.

\clearpage

\begin{figure*}
\vspace{8cm}
\caption{Schematic view of a reconstruction technique which combines images
from different telescopes in a 'common' focal plane viewing part of the sky. 
The scale is in degrees and the simulated images have been recorded from 
7 different telescopes.  The intersection of the image axes measures the
arrival direction of the $\gamma$-ray primary (Courtesy of G.H. Sembroski and M. Kertzman). }
\label{f1}
\end{figure*}

\begin{figure*}
\vspace{8cm}
\caption{Schematic view of a reconstruction technique which combines images
from different telescopes to derive the location of the shower core (where the
shower axis actually intersects with the ground) for a 300 GeV $\gamma$-ray shower.  
Figure shows 7 circles representing the cameras of the 7 telescopes, which are 
located in a coordinate system of the VERITAS telescope array, where the distance
between  telescopes is $\rm \approx$ 85~m.  The individual focal-plane 
also shows the light content in individual pixels (Courtesy of G.H. Sembroski and M. Kertzman).  }
\label{f1}
\end{figure*}

\begin{figure*}
  \vspace{8cm}
\caption{The principle of the detection of Cherenkov light with an
array of heliostats of a solar power plant is shown (STACEE).  Heliostat
mirrors reflect some Cherenkov light onto a secondary mirror of the solar
tower where it is focused onto an array of photomultipliers (Courtesy of R.A. Ong).}
\label{f1}
\end{figure*}

\begin{figure*}
\vspace{8cm}
\caption{The sensitivity of various instruments at GeV to TeV energies.  The values
have been adopted from the following sources: VERITAS~\cite{Vassiliev98}, 
MAGIC~\cite{Barrio98}, GLAST~\cite{Glast96} and Whipple~\cite{Lamb95}.  }
\label{f1}
\end{figure*}



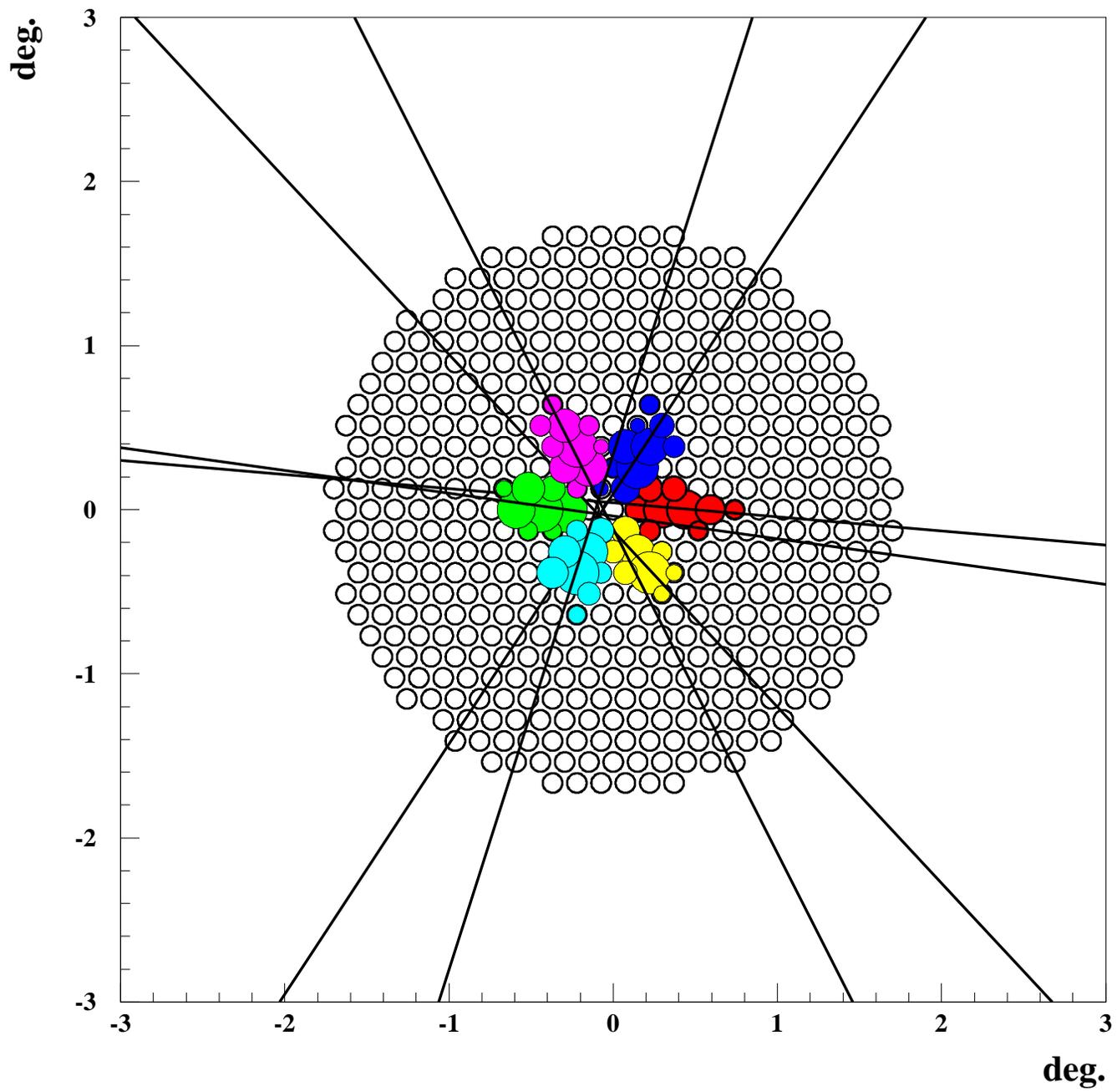
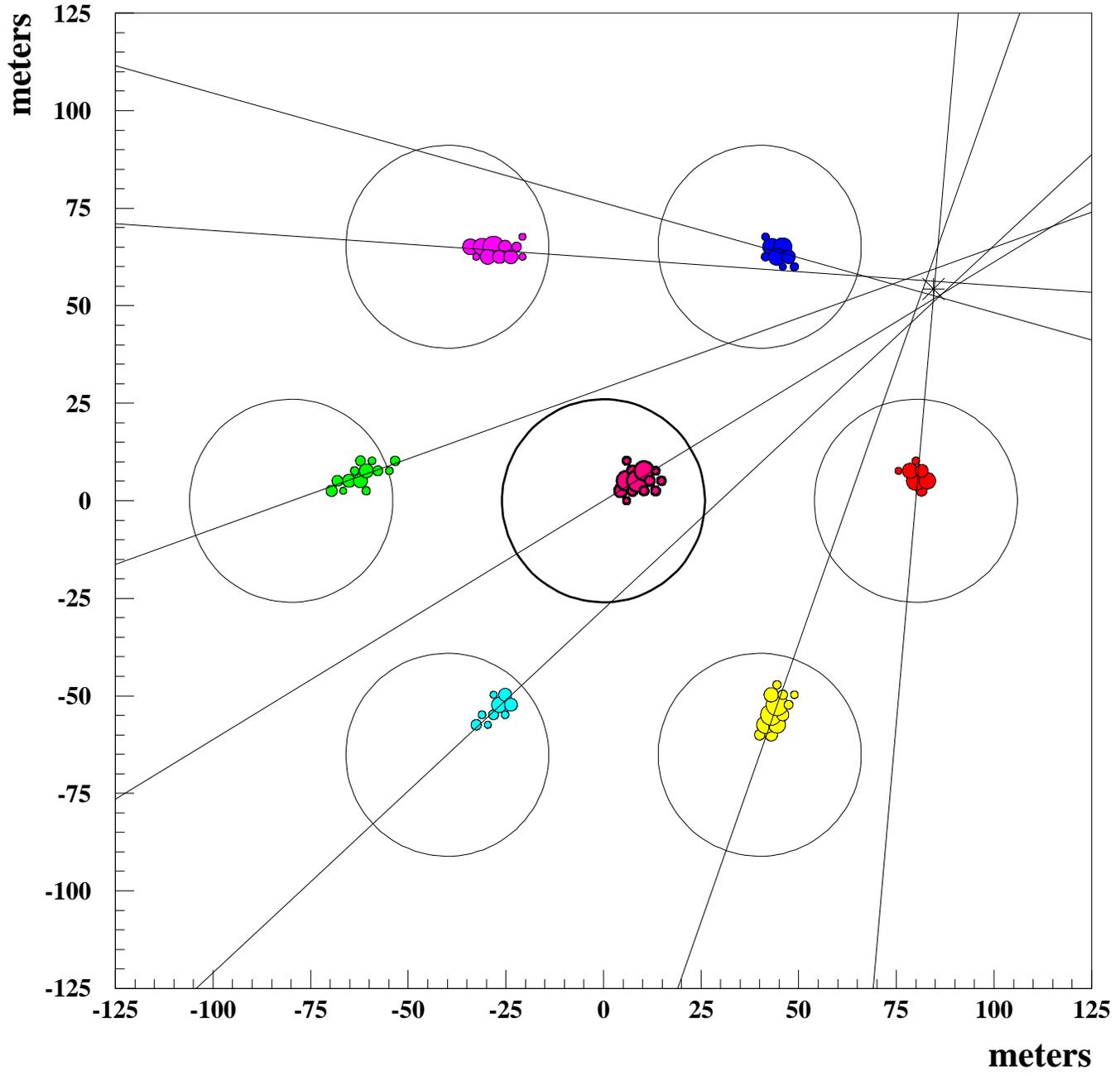
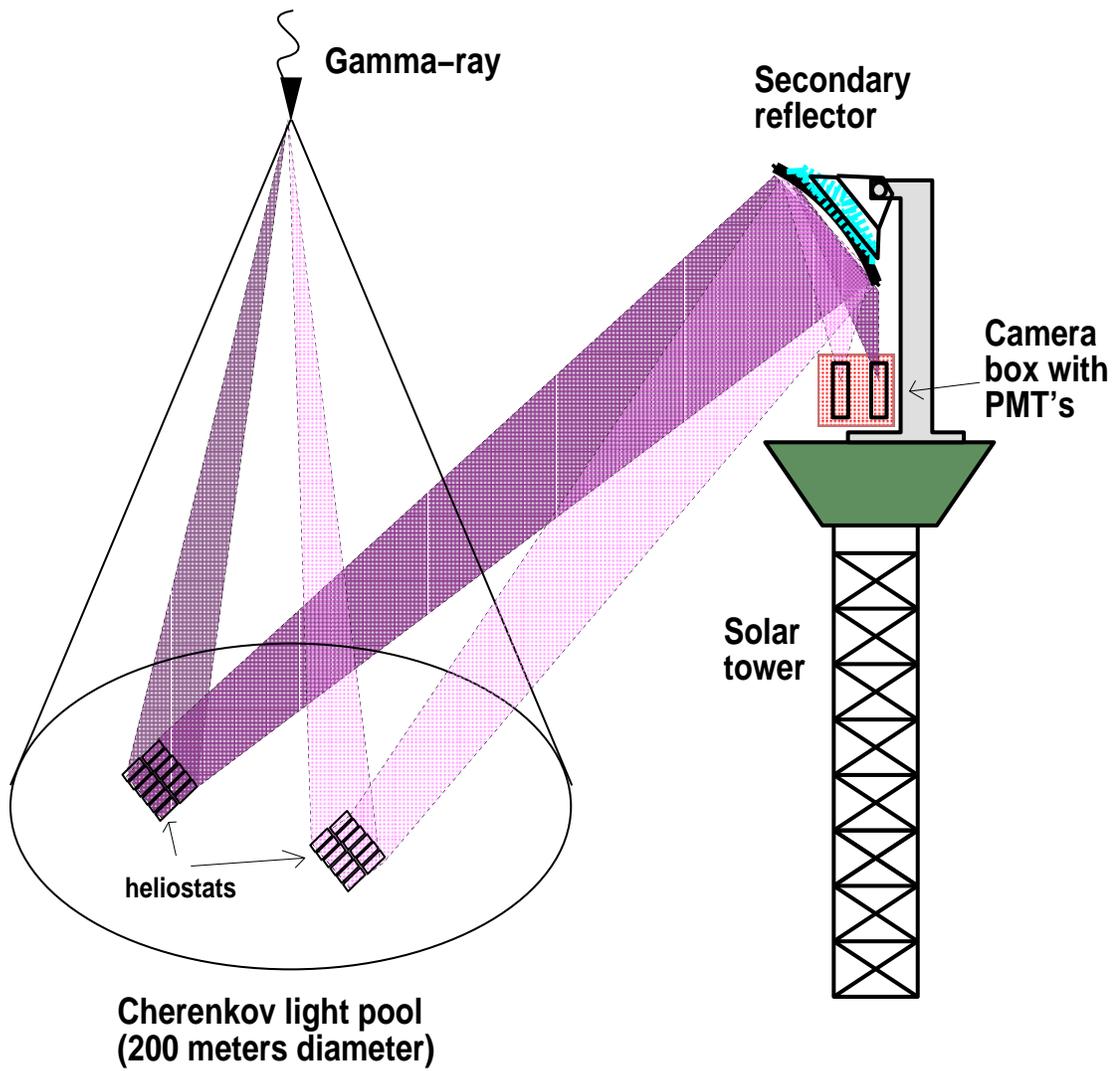
\end{document}